\documentstyle{l-aa}
\input psfig.tex
\def\R{~ROSAT~}
\def\RAS{ROSAT All-Sky Survey }

\def\eta{et al. }
\def\ergs{${\rm erg\,cm^{-2}\,s^{-1}}$ }

\def\mlum{${\rm erg\,s^{-1}\,Hz^{-1}}$ }
\def\mlume{${\rm erg\,s^{-1}\,Hz^{-1}}$}
\def\lo{$l_o$ }
\def\loe{$l_o$}
\def\lx{$l_x$ }

\def\alpox{$\alpha_{ox}$ }
\def\alpoxe{$\alpha_{ox}$}
\def\malpox{$\langle\alpha_{ox}\rangle$ }
\def\malpoxe{$\langle\alpha_{ox}\rangle$}
\def\mGamma{$\langle\Gamma\rangle$ }
\def\mGammaf{$\langle\Gamma_{free}\rangle$ }
\def\mGammag{$\langle\Gamma_{gal}\rangle$ }
\def\z{{\it z} }

\begin{document}
\thesaurus{11.02.1 - 07.01.1 - 17.01.1 - 18.08.1 }
\title{Broad band energy distribution of ROSAT detected quasars
\break II: Radio-quiet objects}
\author{W. Yuan \and W. Brinkmann \and J. Siebert \and W. Voges}
\offprints{W. Yuan}
\institute{Max--Planck--Institut f\"ur Extraterrestrische Physik,
Giessenbachstrasse, D-85740 Garching, FRG}
\date{Received February 8; accepted October 7, 1997}
\maketitle
\markboth{W. Yuan et al.:  ROSAT detected quasars II}{}
\begin{abstract}
A database of radio-quiet quasars\footnote{Table~I is
available in electronic form at the CDS via
anonymous ftp to cdsarc.u-strasbg.fr (130.79.128.5) or via
http://cdsweb.u-strasbg.fr/Abstract.html}
detected with ROSAT is presented
containing 846 quasars seen in
the All-Sky Survey and/or in pointed PSPC observations.
About $\sim $ 70\%  of the objects have been detected in X-rays for the 
first time.
We present the soft X-ray fluxes and spectra, if available.
Using an optically selected subsample compiled from this database
we study  the broad band properties of radio-quiet quasars
with high statistical significance.

We confirm that radio-quiet quasars have in general steeper soft
X-ray spectra (\mGammag = $2.58 \pm 0.05$ for  $z < 0.5$) 
 than radio-loud objects,
with $\Delta\Gamma \sim 0.4 $ and $\Delta \Gamma \sim 0.3$,
 compared to the  flat- and steep-spectrum radio quasars
 (Brinkmann \eta 1997), respectively.
The spectral differences persist to high redshifts
with $\Delta\Gamma \sim 0.6$ at $z > 2$ compared to
flat-spectrum radio-loud quasars.
A spectral flattening with redshift is confirmed
for the radio-quiet objects up to $z \sim 2$,
beyond which the spectral slopes seem to be independent of redshift,
similar to that found for radio-loud quasars.
The spectral slopes of the ROSAT radio-quiet
quasars at $z > 2.5$ ($\Gamma \sim 2.23^{+0.16}_{-0.19}$)
are consistent, within the errors, with
those found for nearby quasars in the medium energy band (2--10~keV).
This implies that X-ray spectral evolution is not important in
radio-quiet quasars.
We show that there is, in a statistical sense,
little or no excess absorption
for most of the radio-quiet objects at $z > 2$,
in contrast to their radio-loud counterparts.

By dividing the sample into narrow redshift bins,  the existence of a
correlation between the X-ray luminosity and the luminosity at
 $2500\AA$, i.e., $l_x \sim l_o^{e}$, is confirmed.
Individual objects show a large scatter from this correlation and
the slope $e$ takes values in the range
 $0.75 \la e \la 1.17$, depending on the mathematical method 
used to analyze the data.

The X-ray loudness \alpox appears to be independent of {\it z}, but
regression analyses indicate 
 a slight increase  of \alpox  with optical luminosity.
However, this  behavior is, very likely, not caused by
physical properties inherent to the quasars but is the result of
the intrinsic dispersion of the luminosities and the flux limits
in both the optical and X-ray observations.

Finally, we find a small fraction of sources with a substantially
larger value of \alpoxe, objects which appear to be  relatively
``X-ray quiet'' compared to the bulk of the other quasars.

\par

\end{abstract}

\section{Introduction}

Convincing evidence for a
correlation between the X-ray and optical luminosity of quasars,
for a surplus of X-ray emission in radio-loud quasars compared to
radio-quiet quasars at the same optical luminosity, and
for differences in the soft X-ray ($\sim$ 0.1--3.5~keV)
spectral properties in the two classes of objects
was provided by {\it Einstein} IPC observations
(c.f. Zamorani \eta 1981, Kriss \& Canizares 1985, Avni \& Tananbaum 1986,
Wilkes \& Elvis 1987, Canizares \& White 1989, Wilkes \eta 1994).
Although the samples used were considerably larger than those of
 previous studies they
did not allow an analysis of the X-ray properties of individual
subgroups of quasars with sufficient statistical significance.
This was especially true for  quasars at higher redshifts,
where the results were ambiguous
because of the small number of detections in X-rays.

The ROSAT observatory (Tr\"umper 1983)
provides for the first time the opportunity
to study a very large number of quasars in the X-ray domain.
More than $\sim$ 30000 AGN are expected to be detected in the
\RAS (RASS) of which the majority are not yet optically identified.
A correlation of source lists from the \RAS and from pointed
observations with existing catalogues
yielded more than 1500 detections of previously known quasars with
many of them seen in X-rays for the first time.
For the radio-quiet quasars this fraction amounts to about 70\%.

In a previous paper (Brinkmann \eta 1997, hereafter paper~I), we have
presented the data for 574 radio-loud quasars\footnote
{We use $\log~R = 1$ as dividing line between radio-loud and radio-quiet
objects, where R is the K-corrected ratio of radio to optical
flux R=f(5GHz)/f(2500\AA). Objects with  $\log~R > 1 $ are called
radio-loud (Stocke \eta 1992, paper~I). We used $\alpha_{radio}$ = 0.5 and
$\alpha_{opt}$ = 0.5 for the K-corrections ($S \propto \nu^{-\alpha}$).}
and 80 quasars detected by ROSAT which show radio emission
but either qualify as ``radio-quiet'' or have no radio-loudness available.
Here we present the second part of 
the ROSAT quasar database---the radio-quiet objects.
We give the  X-ray flux densities and
estimates of the soft X-ray spectral indices, if available,
for objects from the RASS (Voges 1992)
and pointed observations
(ROSAT-SRC, Voges \eta 1995).
We present data  of detected quasars only, although
we will use  upper limits for the non-detections in some places
in the statistical analyses.

In \S~2 we present the data and give details of the
derivation of the relevant parameters,
and we discuss in \S~3 the role of possible X-ray detection biases
of the sample.
The soft X-ray spectra of the radio-quiet quasars are analyzed in \S~4.
We then present the relationship between the X-ray and
optical luminosity in \S~5.
Finally, the conclusions are given in \S~6.

\section{The database of ROSAT detected quasars}

We have compiled a database of all radio-quiet quasars from
the 6th V\'eron-Cetty - V\'eron
quasar catalogue (1993, from now on VV93)
detected in the RASS, as targets of
pointed observations, or as serendipitous sources in pointed
observations available publicly from the
ROSAT-SRC (Voges \eta 1995).
We regard an X-ray source as a detection if the
detection likelihood is equal to or greater than 10 as given by
the Standard Analysis Software System (SASS, Voges \eta 1992).
This threshold corresponds to an about 4$\sigma$ confidence level.
We used an angular distance criterion of $\Delta_{ox} < 60''$ for
the cross correlation as in paper~I.
The total number of ROSAT detected radio-quiet quasars
from the above three sources is 846, of which 289 were seen
in the RASS only, 385 were only seen in pointed observations, and
172 were seen in both the RASS and in pointed observations.
Amongst them are the 69 objects with radio detections, but qualifying
as ``radio-quiet'' according to the above criterion. They 
have already been presented in paper~I.

For all sources from the RASS and pointed observations we use
the results of the SASS employing the most recent
processing of the Survey data (RASS~II, Voges \eta 1996).
For objects which have been seen in both, the \RAS and pointed observations,
we take the data from the pointed observations for the determination of
the spectral parameters and flux densities, as their
statistical errors are considerably smaller due to the generally much longer
exposure in pointed observations.

To estimate the X-ray spectral properties and flux densities
we assume that the soft X-ray spectrum can be represented
by a power law modified by
neutral absorption (see paper~I for details).
For most of the objects the photon indices $\Gamma_{free}$
and the absorbing column densities $N_H$, as well as the photon indices
assuming Galactic absorption $\Gamma_{gal}$ (Dickey \& Lockman 1990)
were estimated
using the two hardness ratios given by the SASS
with the method described in Schartel (1995) and Schartel \eta (1996).
For 134 objects
this method did not yield any reliable spectral parameters at all.
For another 154 objects we could not determine the photon indices $\Gamma_{free}$
and the $N_H$ values simultaneously,  mostly due to the
large statistical fluctuations
in the  small number of photons accumulated from these sources.
Fixing the absorption at the Galactic values yielded, however,  proper
estimates of the spectral indices $\Gamma_{gal}$
of the objects.
On the other hand, for 9 objects with simultaneous estimates
of $\Gamma_{free}$ and $N_H$
no photon indices assuming Galactic absorption, $\Gamma_{gal}$,
could be determined.

The X-ray fluxes in the ROSAT band (0.1-2.4~keV)
were calculated from the count rates using the energy-to-counts
conversion factor (ECF) for a power law
spectrum and Galactic absorption (ROSAT AO-2 technical appendix, 1991).
We used the photon index obtained for an individual source if the estimated
$1~\sigma$ error of $\Gamma_{gal}$ is smaller than 0.5,
and we took the redshift-dependent average
value as derived in \S4.1 below for all objects with larger errors.

For objects with more than one pointed observation,
the mean values of
$\Gamma$, $N_H$, and flux densities as well as their corresponding
errors were calculated, assuming that the source did not vary
between the different measurements.

In Table~1  we list the relevant information for all 846 quasars.
In column 1 we give the IAU designation and, in column 2, a common name.
Objects originally detected at other wave bands than in the optical and UV
(X-ray, infrared, etc.) are marked with stars; a dagger denotes a quasar
found to be radio-loud from recent radio surveys.
In a few cases there is more than one quasar apparently associated with
an X-ray source within $60''$. We list the most plausible object 
(mostly the closest) and mark it with a question mark.
Furthermore, objects for which the X-ray flux is obviously contaminated
by a nearby source (usually extended) are indicated by exclamation marks.
Following the optical positions (J2000), we give
the redshifts and optical magnitudes, as found in VV93.
In column 6 we list the unabsorbed
X-ray flux densities in the 0.1 - 2.4~keV energy band.
The given errors are the statistical $1 \sigma$ errors from the
count rates only. However, for sources with
a small number of counts (mostly from the Survey)
the systematic errors can be of the order of $\sim 30$ \%
(see paper~I).
Further, for strong sources  the spectral fits often show that the
assumed simple power law slope is
 an inappropriate representation of the spectrum.
In both cases the systematic spectral uncertainties
can be considerably larger than the
purely statistical errors and the errors given in Table~1 should,
therefore, be taken as lower limits.
In columns 7 to 10  we give the X-ray  power law photon indices  and the
corresponding $N_H$ values with their $1~\sigma$ errors,
either with free fitted $N_H$  or  obtained
under the assumption of Galactic absorption.
If no error is given, its value is unphysically large.
A missing entry  means that no reasonable spectral index could be obtained.
We used the  Galactic $N_H$ values provided by the EXSAS environment 
(Zimmermann \eta 1994)
which are based on an interpolation of data from Dickey \& Lockman (1990) and
Stark \eta (1992).
In column 11 we indicate whether the object was
detected in the Survey
only (S), in a pointed observation (P), or in both (SP).
If published data are available for an object
we use these results (mostly spectral indices) if they are of superior quality
and we indicate the references in
the last entry (column 12).

\section{The optically selected quasar sample}

In the following we study the broad band energy distribution
of radio-quiet quasars
using a large optically selected sample only.
We compiled this  sample from the above database
by excluding those quasars  which were
 originally found at other wave bands than the optical
(X-ray, infrared, etc.).
We also excluded objects without available redshifts, magnitudes,
or Galactic column densities, and those with uncertain
identifications or with contaminated fluxes.
The optically selected radio-quiet sample thus comprises 644 objects,
of which 202 were seen in the RASS only,
320 were only seen in pointed observations,
and 122 were seen in both the RASS and pointed observations.
It should be noted that the sources detected  in the Survey
 form a well defined
sample as the Survey's limiting sensitivity is rather uniform
(a few times $10^{-13}~{\rm erg\,cm^{-2}\,s^{-1}}$), while the objects
we draw from pointed observations clearly form an
inhomogeneous, incomplete sample because of the  vastly different exposures
and thus different observational sensitivities.

There are nearly 5000 optically selected,
radio-quiet quasars in the VV93 catalogue
not detected either in the RASS or in pointed observations.
As shown in paper~I, the non-detections
in the RASS are mostly due to
the X-ray weakness  of the sources, with fluxes
below the Survey's limiting sensitivity, and only less than 10\%
of the non-detections can be attributed to source intrinsic variability.
Further, some of the non-detections are actually
within a small region of the sky
where the RASS has a relatively short exposure; for example,
about $5\%$ of the non-detections have an exposure less than 100 seconds.
A few objects which are associated with ROSAT sources but at larger
angular distances ($60'' \leq \delta_{ox} < 120''$)
are not regarded as identifications.
We excluded them from the sample of non-detections as well.
Since none of the Broad Absorption Line (BAL) quasars has been detected
by ROSAT (one possible case has been claimed by Green \eta 1995
and Green \& Mathur 1996),
and because there is evidence that BAL quasars  have soft X-ray properties distinct
from the other quasars  (Green \eta 1995, Green \& Mathur 1996),
we excluded them from our analyses.
There are thus about 4000  objects with RASS exposures of more
 than 300 seconds,  with available redshifts, magnitudes,
and Galactic column densities,
which form the subsample of non-detections in our study.

\subsection{Sample characteristics}
Our sample is based on the VV93 quasar catalogue, which
is a compilation of all currently available quasars, and thus
it is heterogeneous and incomplete.
The sample consists of  the cataloged
optically selected quasars without radio detection.
For optically faint objects this does not necessarily mean that they are radio-quiet
as the radio emission of some of them might fall below
the sensitivity limits of the currently available radio surveys.
Using the above dividing line  $\log~R = 1$,
objects fainter than $B \sim 18$ magnitude
(i.e., a large fraction of objects in the sample, see Fig.~1)
will be classified as radio-loud
if their radio flux is greater than $\sim$ 1~mJy---a value 
below all current radio survey flux limits.
In other words, even objects as bright as
$\sim $ 15th magnitude might erroneously be classified
as radio-quiet, because they remain undetected
in the currently most sensitive large scale
radio surveys, i.e. the 87GB survey
($f_{\rm 5GHz}\ga$ 20mJy; Gregory \& Condon 1991)
and the Parkes--MIT--NRAO (PMN) survey
($f_{\rm 5GHz}\ga$ 25 mJy; Gregory et al. 1994)
of the southern hemisphere.
For example, the cross correlation of the VV93 radio-quiet
sources with the PMN radio survey yielded 55 objects,
which are thus radio-loud; 5 of them were detected
by ROSAT (marked with daggers in Table~1) and
excluded from the analysis. Similarly treated were 12 objects from a
recent quasar correlation with the NVSS survey (Bischof \& Becker 1997).
However, we still expect most of the quasars without radio detection
to be radio-quiet because of the generally
low fraction of radio-loud objects
($\sim 15\%$, Kellermann \eta 1989) among optically selected
quasars.

\subsection{Detection biases}
We plot in Fig.~1 the cumulative number-magnitude diagram
for the whole optically selected, radio-quiet quasar sample
(squares), for all the objects detected by ROSAT
in the RASS or in pointed observations (triangles),
and those seen in the RASS only(circles).
For illustrative purpose we show as a dashed line the slope
for uniformly populated sources in Euclidean space,
normalized at magnitude $B = 16$.
The curve for the RASS detected sources flattens already
at $B \sim 16$ due to the limited sensitivity of the
X-ray survey, which results in a lower X-ray detection
rate at the faint end of the magnitude distribution
compared to the total sample.
The inclusion of objects detected in pointed observations
greatly enlarges the size of the sample of X-ray detected quasars,
especially at fainter optical magnitudes.
However, it does not improve substantially the completeness
of the sample of X-ray detections as the accumulated area of the pointed
observations covers only $\sim 10\%$ of the sky.

\begin{figure}
\psfig{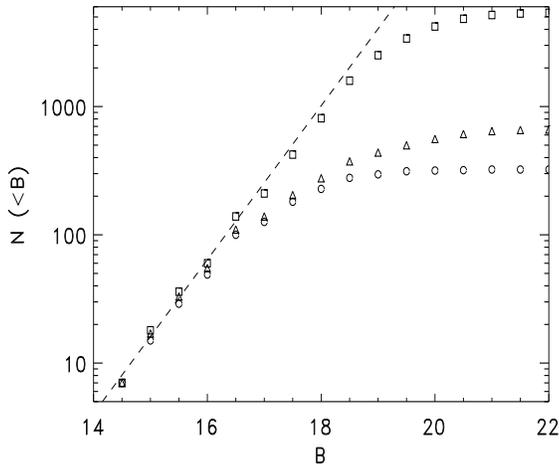}
\caption[]{The number-magnitude diagram for
the whole optically selected, presumably radio-quiet sample
in the VV93 quasar catalogue (squares),
the ROSAT detections from the RASS or pointed observations (triangles)
and the objects detected in the RASS only (circles).
The dashed line is the slope (normalized at $B = 16$)
for uniformly populated sources in Euclidean space
for illustrative purpose.}
\end{figure}

The \RAS has a relatively uniform limiting sensitivity of a
few $\times ~10^{-13}$ \ergs with the exact value depending slightly
on the amount of intervening Galactic absorption,
on the exact shape of the X-ray spectrum of the source,
and on the local Survey exposure.
In Fig.~2 we plot the detection rate (in percent) of the RASS for the
radio-quiet quasars (hatched) as a function of magnitude,
i.e., the number of objects detected in the
RASS in a magnitude bin divided by the total number of
objects at that magnitude in VV93.
Objects with exposures in the RASS of less than 300 seconds
 are excluded.
For comparison the detection rate for the radio selected
radio-loud sample (see paper~I) is also presented.
Apart from the flux limit of the Survey,
the detection probability at a given magnitude depends on
the ratio of the flux densities in the X-ray and optical wave bands.
Beyond 16~mag, the continuous decrease of the detection rate
with increasing $B$
over about 3 magnitudes suggests a substantial dispersion of
the optical flux density (a factor of $\sim 16$) for the RASS objects
with an X-ray flux at the Survey limit,
and thus, a large dispersion in
the ratio of the X-ray-to-optical flux densities for radio-quiet quasars.
This implies a similar dispersion of the luminosity ratios
 between the X-ray and optical
wave bands, the X-ray loudness, since the effect of different
K-corrections in the two wave bands is small
(a factor of $1+z$ for the typical optical and X-ray spectral indices
of $\alpha = 0.5$ and $1.5$, respectively, $S \propto \nu^{-\alpha}$).
We will study the X-ray-to-optical luminosity ratio
quantitatively in detail in \S~5.
It is noted that even at very bright optical magnitudes ($B \sim 15$)
there exist a number of objects which remain undetected
in the RASS and which must thus have very low X-ray-to-optical flux ratios.
Some of them, detected later in pointed observations,
show unusually weak soft X-ray fluxes
relative to their optical emission
compared to the bulk of the optically selected quasars, and
will be discussed in \S5 below.

Large differences in the detection probabilities can be seen
between the radio-loud and radio-quiet quasars.
The radio-loud quasars have X-ray detection
rates significantly higher than radio-quiet objects at a given
optical brightness,
and they are  detected at much fainter optical magnitudes
(down to $\sim 21$~mag).
Given the average X-ray-to-optical luminosity ratio
found previously (Wilkes \eta 1994, Green \eta 1995, paper~I)
radio-loud quasars are X-ray brighter
than radio-quiet objects by a factor of $\sim 3 - 4$.
If we reduce artificially the measured X-ray luminosities of
the radio-loud objects by this factor,
taking into account the appropriate K-corrections,
we find that many of them would not have been detected in the RASS
and the detection probabilities of
the radio-loud and radio-quiet quasars would then roughly match.
This means that the different detection probabilities are mainly caused by
the differences in the X-ray-to-optical luminosity ratios
between the two classes.

\begin{figure}
\psfig{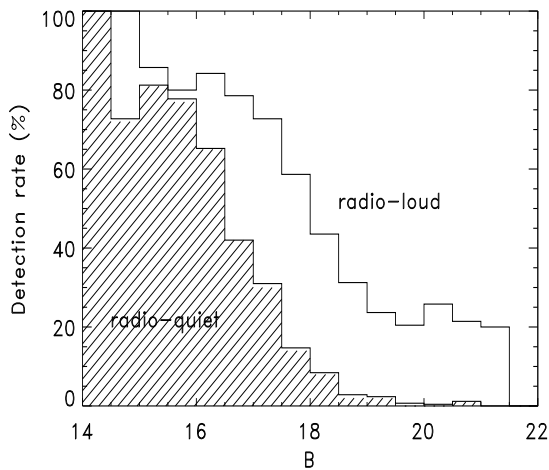}
\caption[]{Detection rate in percent of quasars in the RASS as
a function of optical magnitude for both radio-quiet (hatched)
and radio-loud (open) quasars.}
\end{figure}

\begin{figure}
\psfig{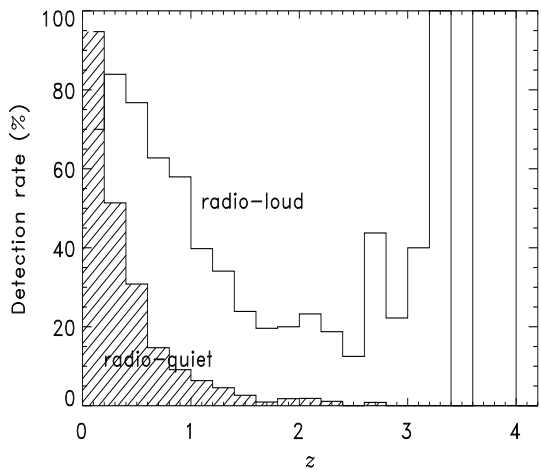}
\caption[]{Detection rate in percent of quasars detected in the RASS
as a function of redshift for both radio-quiet (hatched)
and radio-loud (open) quasars.}
\end{figure}

In Fig.~3 the  detection rate
is shown  as a function of redshift.
The most distant radio-quiet quasar detected in the RASS
is 0053-2532 at redshift $z = 2.70$, however, with a relatively large
angular distance of $\sim 50''$ between the X-ray and optical position
and a high X-ray-to-optical luminosity ratio of
$\alpha_{ox} \sim 1.0$ (see \S5 for $\alpha_{ox}$).
The X-ray detection rate of radio-quiet quasars drops much
faster towards higher $z$ than that of radio-loud objects,
and no sources are found in the RASS at  $z > 3$.
In contrast, radio-loud quasars were seen in the RASS
up to very high redshifts ($z \sim 4$),
and the detection rate appears to increase towards higher redshifts.
However, this increase
might not be statistically significant as the total number of
radio-loud quasars at z $\geq 2.5$  is 48 of which 20 were
detected (cf. Paper 1).

In Fig.~4 we show the K-corrected broad band soft X-ray
luminosities (0.1--2.4~keV) of
the sources in our sample as a function of redshift
(see \S~5 for details of the calculation of the luminosities).
We have used a Friedman cosmology with
$H_0 = 50~{\rm kms^{-1}Mpc^{-1}}$ and $q_0 = 0.5$
for the  computation of the luminosities.
For illustrative purposes we plot (full curve)
the K-corrected luminosity corresponding to the
typical Survey detection limit of 
$4 \times 10^{-13}$ \ergs.
It shows that many sources have luminosities definitely below the
Survey's sensitivity
 limit and they could thus be detected only in pointed observations.

\begin{figure}
\psfig{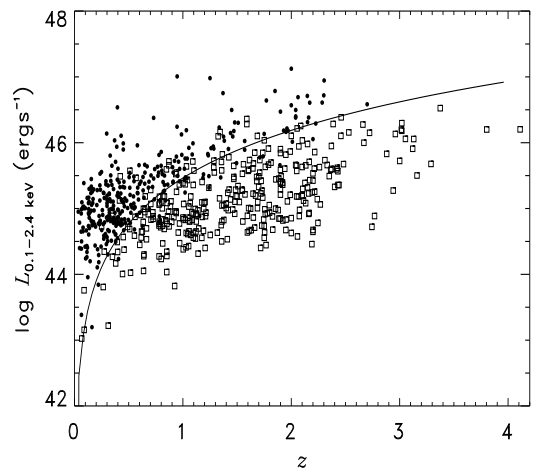}
\caption[]{Integrated X-ray luminosity (0.1--2.4~keV) as a function of
 redshift for the radio-quiet quasar sample.
Filled circles are sources seen in the RASS, and open squares
are objects detected in pointed observations only.
The full curve represents the K-corrected luminosity
of a source at the typical Survey flux limit of $4 \times 10^{-13}$~{\ergs}.}
\end{figure}

\subsection{Variability}
More than 100 quasars have been observed repeatedly in pointed
observations and they thus provide a good sample for an evaluation
of the X-ray variability of quasars. In Fig.~5 we plot a histogram
of the observed maximal variability of the objects, i.e., the ratio between
the highest and the lowest count rates.
The distribution of the count rate ratio follows approximately
a Gaussian with a $\sigma \sim 0.5$,
which is indicated by the dashed curve in the figure.
Similar to the radio-loud sample,
Fig.~5 shows that a large fraction of the quasar
population is variable, mostly by less than a factor of two in flux.
For weak sources low intrinsic variability cannot be distinguished from
statistical fluctuations.
No extremely variable quasars (by a factor of 5 or more),
as found in the radio-loud sample
in paper~I, seem to exist.
However, if we compare the X-ray fluxes of sources observed in
pointed observations with the corresponding RASS fluxes, for which the
uncertainties are larger, we do find one case of extreme flux variation
by a factor of six (PG 0844+349)
with a count rate of $\sim 0.568 \pm 0.04$ (413~s exposure) in the RASS and
$\sim 0.095 \pm 0.003$ in the pointed observation, respectively.
Interestingly, no statistically significant spectral changes were found between
the two observations and the X-ray spectrum is not particularly steep
($\Gamma \sim $2.5 $\pm$ 0.1).
Unfortunately, as most of the repeated observations of a source
have been performed in different \R pointing periods
we cannot make any reliable estimates about the
variability time scales and the
relation between the time scale and variability amplitude of a source.

\begin{figure}
\psfig{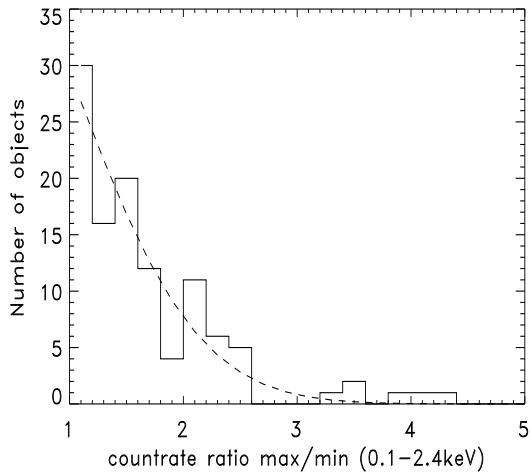}
\caption[]{Histogram of the variability of quasars seen more than
 once in pointed observations. The ratio of the
 highest and the lowest observed count rates is used as a
 measure of flux variability.
 The dashed curve is the best fit of a  Gaussian distribution.}
\end{figure}

\section{X-ray spectra of radio-quiet quasars}

\subsection{Mean spectral indices}
Spectral studies of radio-quiet quasars in the medium energy range
 yielded power law slopes
\mGamma $= 1.90 \pm 0.11, \sigma = 0.20^{+0.16}_{-0.12}$
in the 2--10~keV band observed by {\it EXOSAT} (Lawson \eta 1992)
and \mGamma $= 2.03 \pm 0.16, \sigma = 0.13^{+0.17}_{-0.06}$ (90\% confidence)
in the 2--20~keV band observed by {\it Ginga} (Williams \eta 1992).
In the softer energy range quasars show a wide variety of spectral indices with
a mean \mGamma$\sim 2$ (0.2--3.5~keV) from {\it Einstein} IPC observations
(Wilkes \& Elvis 1987, Canizares \& White 1989).
In the even softer ROSAT PSPC energy band (0.1--2.4~keV),
spectral studies of quasars
generally give steeper spectra ($\Delta \Gamma \sim 0.5$)
and a wider dispersion of the indices
than found in the harder energy band
(Brinkmann 1992, Brunner \eta 1992, Fiore \eta 1994, Schartel \eta 1996,
Laor \eta 1997).
This steepening of the power law slopes is commonly explained
by an additional soft component in the ROSAT PSPC bandpass
(see Mushotzky \eta 1993 for a review),
similar to that seen in Seyfert~I galaxies
(Turner \& Pounds 1989, Masnou \eta 1992, Walter \& Fink 1993).
Radio-loud quasars are found to have systematically
flatter spectral indices than radio-quiet objects
over the total X-ray range, by
$\Delta \Gamma \sim 0.3$ in the 2--10~keV energy band
(Lawson \eta 1992, Williams \eta 1992),
$\Delta \Gamma \sim 0.5$ in the 0.2--3.5~keV band
(Wilkes \& Elvis 1987, Canizares \& White 1989), and
$\Delta \Gamma \sim 0.2 - 0.3$ in the 0.1--2.4 keV band
(Brunner \eta 1992, Schartel \eta 1996, paper~I).
An extra hard X-ray component which is linked to the
radio emission of radio-loud quasars might be an explanation for
this differences in spectral indices.

We used a maximum-likelihood  method
described in Maccacaro \eta (1988) and Worrall \& Wilkes (1990)
to estimate the intrinsic distribution of the power law photon indices,
assuming that both the distribution of the indices and
the uncertainties of the measurements are Gaussian.
To minimize possible redshift effects, we split up the sample
into different redshift ranges, as defined in Table~2.
We considered only objects for which both indices,
$\Gamma_{free}$ and $\Gamma_{gal}$, are available.
For a few objects with extremely small errors
of the measured spectral indices due to
very high photon statistics
we use  a lower bound for the systematic uncertainties of
$\delta \Gamma = 0.05$ to avoid the large statistical weight
of individual strong sources.
Further, in many cases  
a simple power law is usually not a good representation
of the soft X-ray spectrum for these objects.
The best estimates of the mean and standard deviation
(errors are at joint 68\% confidence level for two parameters)
are listed in Table~2.
We plot the best estimates (crosses) and the 90\% confidence contours
in Fig.~6 for objects at both low ($z < 0.5$) and high ($z > 2$) redshifts,
for the power law photon indices $\Gamma_{free}$ obtained with
free $N_H$ (solid line) and for the indices $\Gamma_{gal}$ obtained
with fixed Galactic $N_H$ (dashed line), respectively.

\begin{figure}
\psfig{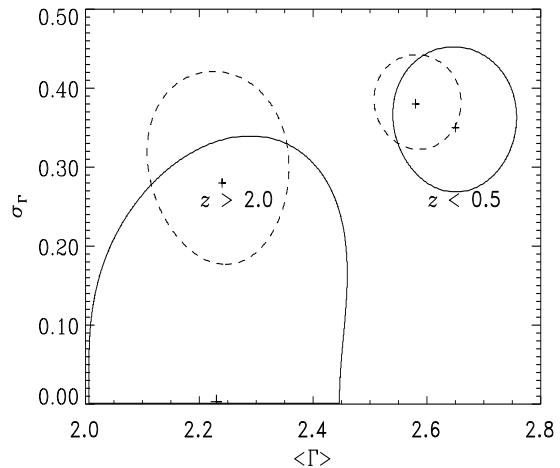}
\caption[]{Best estimates (crosses) and 90\% confidence contours of the
mean \mGamma and dispersion $\sigma$ of the photon index
distribution for the $z < 0.5 $ and $z > 2$ objects.
Solid curve: \mGammaf obtained with free absorption;
dashed curve: \mGammag obtained with absorption fixed at the Galactic
column density.}
\end{figure}

\begin{table*}
\setcounter{table}{1}
\caption[]{Mean spectral photon index and dispersion in different redshift bins}
    \begin{flushleft} \begin{tabular}{lccccc}
    \hline\hline\noalign{\smallskip}
 redshift & number  &  \mGammaf & $\sigma_{free}$ & \mGammag & $\sigma_{gal}$  \\
 \noalign{\smallskip}  \hline \noalign{\smallskip}
      $ z < 0.5$          & 146 & $2.65^{+0.08}_{-0.08}$ & $0.35^{+0.07}_{-0.06}$ & $2.58^{+0.05}_{-0.05}$ & $0.38^{+0.04}_{-0.04}$ \\
      $ 0.5 \leq z < 1.0$ & 81  & $2.54^{+0.16}_{-0.16}$ & $0.31^{+0.16}_{-0.16}$ & $2.46^{+0.07}_{-0.07}$ & $0.34^{+0.06}_{-0.05}$ \\
      $ 1.0 \leq z < 2.0$ & 107 & $2.36^{+0.13}_{-0.14}$ & $0.20^{+0.16}_{-0.15}$ & $2.35^{+0.06}_{-0.06}$ & $0.34^{+0.06}_{-0.05}$ \\
      $ z > 2.0$          &  56 & $2.23^{+0.15}_{-0.15}$ & $0.00^{+0.20}_{-0.00}$ & $2.22^{+0.09}_{-0.09}$ & $0.28^{+0.10}_{-0.08}$ \\
      $ z > 2.5$          &  15 & $2.20^{+0.53}_{-0.53}$ & $0.00^{+0.61}_{-0.00}$  & $2.23^{+0.16}_{-0.19}$ & $0.26^{+0.20}_{-0.12}$ \\
    \noalign{\smallskip}  \hline
    \end{tabular}  \end{flushleft}
Note: errors are at joint 68\% confidence level for 2 parameters
\end{table*}

The two mean spectral photon indices  \mGammaf and
\mGammag are compatible with each other within their 1~$\sigma$ errors,
implying a general agreement between the value of
the absorbing column density determined
from fitting the X-ray spectra and the Galactic value
from HI radio measurements at all redshifts.
For objects at $z < 0.5$ the
mean photon index (\mGammag $= 2.58 \pm 0.05$,
$\sigma_{gal} = 0.38^{+0.04}_{-0.04}$)
is in good agreement with that found by Schartel \eta
(1996; \mGammag $=2.54\pm0.04$, $\sigma_{gal}=0.24\pm0.03$).

At high redshifts ($z \ga 2.5$), the energy band of the X-ray spectrum
measured with ROSAT corresponds to a rest frame energy of up
to $\sim$~10~keV,
similar to the {\it EXOSAT/Ginga} bandpass for observations of
low redshift objects.
We found  \mGammag$= 2.23^{+0.16}_{-0.19}$ and
$\sigma_{gal} = 0.26^{+0.20}_{-0.12}$, in good agreement with
Bechtold \eta (1994b) who give \mGammag$= 2.15 \pm 0.14$ using a sample
of 6 objects from ROSAT pointed observations.
The mean spectral slope and dispersion at $z \ga 2.5$ are consistent,
within $\sim 1\sigma$ errors, with those found for
low redshift quasars by {\it EXOSAT}
(\mGamma $= 1.90 \pm 0.11, \sigma = 0.20^{+0.16}_{-0.12}$; Lawson \eta 1992) and
{\it Ginga} (\mGamma $= 2.03 \pm 0.16,
\sigma = 0.13^{+0.17}_{-0.06}$, 90\% confidence; Williams \eta 1992).
The marginal difference of the mean spectral indices
could result from the lower bound of the source intrinsic energy range
in the ROSAT observations ($\sim$ 0.4 keV for quasars at {\it z} = 2.5,
 compared to
$\sim $ 2 keV for EXOSAT and {\it Ginga} for local quasars).
Thus, the spectral slopes in the intrinsic energy range up to 10~keV
for high and low redshift quasars are comparable,
which implies that X-ray spectral evolution is
not important in radio-quiet quasars.
The steep spectra of radio-quiet quasars seen at high redshifts
strengthen the claims that
they cannot be the dominant contributors
to the cosmic X-ray background,
unless their spectra flatten substantially at higher energies
(Fabian \& Barcons 1992).
However, they may have a considerable contribution to the
possible ``soft excess'' of the cosmic X-ray background
at softer energies (Hasinger \eta 1993, Gendreau \eta 1994).

The results confirm the long established difference of
spectral indices between the radio-loud and radio-quiet quasars at low redshifts
(Wilkes \& Elvis 1987, Canizares \& White 1989, Brunner \eta 1992).
Radio-quiet quasars show steeper power law spectra by
$\Delta\Gamma \sim 0.4 $ and $\Delta\Gamma \sim 0.3$ compared to the
flat- and steep-spectrum radio quasars (paper~I), respectively.
This  difference persists and gets even larger at redshifts $z > 2$
by $\Delta~\Gamma \sim 0.6$.

\subsection{Redshift dependence of spectral indices}
ROSAT studies of quasars  revealed
a flattening of the spectral slope with increasing redshift
both for radio-loud (Schartel \eta 1992, Schartel \eta 1996, paper~I)
and for radio-quiet objects (Stewart \eta 1994).
This trend is found to disappear at redshifts higher than $z \sim 2$
for the radio-loud quasars (paper~I).
The spectral indices at these redshifts,
for which the observed energy band corresponds to
$\sim$~0.3--7~keV in the quasar's rest frame,
are consistent with those found for nearby objects in the medium
energy band by {\it EXOSAT} and  {\it Ginga}.
The aforementioned composite continuum spectrum of quasars
(Stewart \eta 1994, Schartel \eta 1996),
which consists of a hard component seen in the medium energy band
and a steep component at softer energies
can account for this effect in that
the soft component moves out of the ROSAT energy band
with increasing redshift.
This photon index -- redshift dependence has been modeled by
Schartel \eta (1996) by fitting a two-component spectrum
to the data  of radio-loud quasars.

It can be seen in Table~2 that
the mean photon index flattens progressively with increasing redshift.
In Fig.~7 we plot $\Gamma_{gal}$ as a function of redshift for a subsample
of 488 quasars for which the determined photon indices
$\Gamma_{gal}$ have errors of $\delta \Gamma_{gal}< 1.0$.
A Spearman rank correlation test 
yields a correlation coefficient $R_s = -0.29$ (N=488) which corresponds
to a probability for such a correlation to occur by chance of 
 $P_r \simeq 3 \times 10^{-11}$ for the whole subsample,
and a probability of $P_r \simeq 5 \times 10^{-6}$ for the $z \leq 2$ objects,
indicating a correlation with redshift at a high significance level.
However, the correlation weakens significantly and even disappears at higher redshifts:
for $z > 2$ we find $R_s = -0.09$ (N=70), $P_r = 0.45$, and
for $z > 2.5$, $R_s = 0.15$ (N=17), $P_r = 0.58$, respectively.
In the $z \leq 2$ range a regression analysis
without weighting yielded the relation
\mGammag$= 2.63(\pm 0.04) - 0.20(\pm 0.04) \times z $,
as indicated by the dashed line in Fig.~7.
At $z > 2$ the mean \mGammag is $\sim 2.22 \pm 0.09$ (see Table~2, 
the conventional mean without weighting is $\sim 2.19$),
which is indicated by the horizontal dashed line in the figure.
The choice of the break redshift in the above analysis is somewhat
arbitrary, but it seems to be around $z \sim 2$.
Furthermore, in  Table~2,
there seems to be a progressive decrease in the
dispersion $\sigma_{gal}$ of the intrinsic distributions of spectral slopes
with increasing redshift,
though all values are consistent within their $1~\sigma$ errors.

\begin{figure}
\psfig{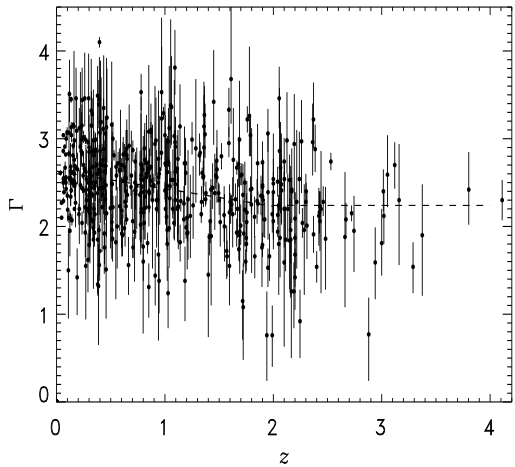}
\caption[]{X-ray photon index $\Gamma_{gal}$ as a function of redshift
for radio-quiet quasars detected by ROSAT.
The resulting linear regression line is indicated as a dashed line (see the text).}
\end{figure}

We thus confirm a flattening of the soft X-ray spectra with
increasing redshift for our large sample of  radio-quiet quasars.
This trend is found to vanish beyond a redshift of $\sim 2$,
similar to that found for the radio-loud quasars (paper~I).
The continuity of the spectral flattening with redshift,
as seen in Table~2 and Fig.~7,
supports the two-component model of the soft X-ray spectrum of quasars.
The wide range of the power-law indices implies that
the soft excess component varies from object to object
and in  time, which  is supported by observations of individual objects 
 (Elvis \eta 1991, Masnou \eta 1992).
The cutoff at redshift of $\sim 2$ suggests that the soft excess contribution
to the X-ray continuum
becomes negligible above $\sim 0.3$~keV in the quasar's rest frame.
The similarity in the spectral index--redshift relation between radio-loud and
radio-quiet quasars suggests a similar origin of
the soft X-ray emission in these two object classes.

\subsection{Absorbing column density}

Previous studies of quasars with the {\it Einstein} IPC gave no indications
for intrinsic excess absorption
(Canizares \& White 1989, Wilkes \& Elvis 1987).
For low redshift quasars, ROSAT observations confirmed these results
showing good agreement between the absorbing column densities
obtained from spectral fits and those inferred from HI observations
(Schartel \eta 1996, Laor \eta 1997).
Recent analyses of ROSAT and ASCA observations of
high-$z$ ($z \ga 3$) radio-loud quasars reveal, however,
that excess absorption is common in these objects
(Elvis \eta 1994, Siebert \eta 1996, Cappi \eta 1997, and paper~I)
and might even be temporarily variable (Schartel \eta 1997).
Radio-quiet quasars probably
do not show this effect (Bechtold \eta 1994a), which indicates that
this excess absorption is intrinsic to the radio-loud quasars, but
the statistics on which this argument is based is poor.
The search for absorption in radio-quiet quasars is hampered
by the fact that for most of the high-$z$
objects observed by \R not enough photons could
be accumulated for a precise
determination of the spectral parameters for the individual object.
Furthermore, even if there exists excess absorption
it will mainly effect the lower part of the spectral energy band which
will be redshifted out of the \R bandpass if the absorber
is at high redshifts as well.
This usually results in
large uncertainties in the determination of the absorbing column density.
However, the large number of objects at high redshifts available in
Table~1 allows a statistical analysis of the presence of this
excess absorption in our sample.

For a statistical test the mean value of the differences
between the fitted  and the Galactic column densities,
 $\Delta N_H = N_H - N_{H,gal}$, is difficult to evaluate
because the  fitted $N_H$  values are
asymmetrically distributed with respect to the actual $N_{H,gal}$,
with a longer tail towards higher $N_H$ values.
For example, the simple method to calculate the mean by
assuming a normal distribution usually gives an overestimate for the mean.
This problem is even more severe in the case of
data with poor statistical quality, which results in
large uncertainties in the deduced $N_H$ values.

We used an indirect, yet simple, non-parametric statistics
to test for the presence of excess absorption.
We assume that there is no systematic discrepancy between
the X-ray measured column densities and the
radio HI results, and that there are no systematic errors
in the measurements.
If there were no intrinsic absorption in any  quasar
the distribution of $\Delta N_H$ should have
a mean of zero and a median not greater than the mean.
For a sample in which the data are obtained with
high statistical significance,
the distribution is approximately symmetric
and the median would be close to the mean.
Then, if the hypothesis of ``no excess absorption'' is true,
the median of $\Delta N_H$ should not be  greater than zero,
or, equivalently, the number of objects with $\Delta N_H \leq 0$
(denoted as $\Sigma$) should not be smaller than half of the total sample,
with scatter caused by the uncertainties of the X-ray measurements.
We use the quantity $\Sigma$ as a test statistic (binomial test, one-tail)
to test the hypothesis.
We ignored objects
for which the corresponding $N_{H,gal}$ values are higher than
$5 \times 10^{20}~{\rm cm^{-2}}$ as at these values
the fitted spectral parameters tend to have
large uncertainties in the PSPC energy band.
We also ignored objects with large errors in the fitted absorption column
($\delta N_H > 2 \times 10^{21}~{\rm cm^{-2}}$).
This yielded 340 objects for which $N_H$ values could be fitted,
which were further divided into four redshift bins, as defined in Table~3.
The number of objects in each bin, the median of the
normalized $\Delta N_H/N_{H,gal}$, the $\Sigma$ value,
and the corresponding probability levels $P$
that  the corresponding subsample is drawn from a parent population
without excess absorption
are listed in the upper panel of Table~3.

\begin{table}
\caption{Results of the binomial test for no excess absorption}
    \begin{flushleft} \begin{tabular}{lrcrl}
    \hline\hline\noalign{\smallskip}
 subsample & number  &   median & $\Sigma^{a)}$ & $P^{b)}$ \\
 \noalign{\smallskip}  \hline \noalign{\smallskip}
      $ z < 0.2$          & 53  &  0.01 & 25  & 0.34 \\
      $0.2 \leq z < 1.0$  & 142 &  0.14 & 62  & 0.06 \\
      $1.0 \leq z < 2.0$  & 98  &  0.22 & 40  & 0.04 \\
      $  z \ge 2.0       $  & 47  &  0.40 & 15  & 0.006 \\
    \noalign{\smallskip}  \hline
    \multicolumn{5}{c}{objects with $> 100$ photons} \\
    \noalign{\smallskip}  \hline \noalign{\smallskip}
      $ z < 0.2$          & 45  &  0.01 & 21  & 0.33 \\
      $0.2 \leq z < 1.0$  & 79  &  0.03 & 39  & 0.45 \\
      $1.0 \leq z < 2.0$  & 41  &  0.01 & 19  & 0.32 \\
      $  z \ge 2.0     $  & 13  &  0.16 &  5  & 0.23 \\
    \noalign{\smallskip}  \hline
    \end{tabular}  \end{flushleft}
a) number of objects with $\Delta N_H \leq 0$   \\
b) the probability level that the sample 
is drawn from a parent population without excess absorption
\end{table}

The data are consistent with no or little excess absorption at low redshifts,
in agreement with what has been found in previous studies.
At high redshifts ($z \ge 2$) the data seem to suggest a surplus
in the column densities obtained from the X-ray data,
though the significance is not very high.
We have checked the possibility that the differences result from
different $N_{H,gal}$ distributions  for the objects
at high ($z \ge 2$) and low redshifts ($z < 0.2$).
The two $N_{H,gal}$ distributions are consistent with
each other ($\chi^2$ test, with a reduced $\chi^2 = 0.87$ for 9 d.o.f).
Further, no correlation is found between the obtained
$\Delta N_H$ and $N_{H,gal}$ in our data.
A plausible explanation might be the
low photon statistics in some of the high-redshift quasars
for which only relatively few photons could be accumulated.

We thus restrict our analysis to objects with more than 100
source counts and list the result in the lower panel of Table~3.
For $z \ge 2$ objects the probability level is reduced to 0.23 only,
and the data are consistent with no excess absorption.
The previously derived $P$ including objects with low  numbers of
source photons is thus, at least partly, an artifact of systematic biases.
Unfortunately, the small size of the sample
(with  more than 100 source photons) makes such a test rather insensitive
when the fraction of objects with possible ``excess absorption''
is small.
More observational data for high-$z$ radio-quiet objects are needed
in order to obtain results with higher statistical significance.

We conclude that the current data are, in a statistical sense,
consistent with no or only little excess absorption
at all redshifts for radio-quiet quasars,
in contrast to  their radio-loud counterparts,
which show a high incidence of excess absorption at higher redshifts.
At redshifts beyond $z = 2$
more observational data are needed for radio-quiet quasars.

\section{Broad band energy distribution}

The knowledge of the exact form of the relation between the X-ray and optical
luminosities is required to relate the quasar
statistics (evolution, luminosity function) in the two wave bands
and to understand the quasars' broad band emission.
The X-ray-to-optical energy distribution is commonly characterized
by $\alpha_{ox}$, the broad band spectral index from
the ultra-violet ($2500\AA$) to the X-ray (2~keV) which is defined 
as the luminosity\footnote{In the following, we will for convenience refer to the
luminosity $l_{2500{\AA}}$ as to the ``optical luminosity'', $l_o$.}
  ratio
$ \alpha_{ox} = - 0.384~\log~(l_{\sf2keV}/l_{\sf2500{\AA}}$).
A dependence of the X-ray luminosity of the form  $l_x \sim l_o^e$
is equivalent to the relation
$\alpha_{ox} \sim \beta\log~l_o$ with $e = 1 - 2.605 \times \beta$.

Previous studies of optically
selected quasars gave the following results:
$\alpha_{ox}$ is nearly independent of redshift but
it increases with $l_o$, or equivalently,
there exists a non-linear relationship
$l_x \sim l_o^e$ with $e \sim 0.7 - 0.8$
(Kriss \& Canizares 1985, Avni \& Tananbaum 1986,
Wilkes \eta 1994, Avni \eta 1995).
Using ROSAT survey data for the Large Bright Quasar Survey sample,
Green \eta (1995) found a similar non-linear relationship.
For models of a pure luminosity evolution of quasars,
this implies a slower luminosity evolution in X-rays than in the optical.
Interestingly, there are recent reports suggesting a linear relationship
$l_x \sim l_o$ and, consequently, \alpox being independent of $\l_o$
(La Franca \eta 1995). The results were achieved by applying a
regression analysis to the {\it Einstein} data which
takes into account the errors in both variables.

Using the large sample of ROSAT detected radio-loud quasars
and similar regression methods, Brinkmann \eta (paper~I) also found a
linear relationship $l_x \sim l_o$ for a subsample
($29.6 < \log~l_o < 31.7$)
of flat-spectrum quasars and a marginally flatter
slope, $e \sim 0.94 \pm 0.13$, for the whole sample of both
flat- and steep-spectrum quasars.
But the existence of an X-ray component related to the
radio emission in radio-loud quasars
introduces additional complications in the analysis.
Radio-quiet quasars, therefore, seem to be better suited to study
the relationship between the optical and X-ray emission,
which is believed to originate from the active nucleus.

We convert the observed X-ray fluxes into luminosities as in paper~I.
For the K-correction and the calculation of the monochromatic X-ray luminosity,
the individual spectral index for an object is used
if it is obtained with relatively small error ($\delta~\Gamma < 0.5$),
otherwise the redshift dependent mean spectral index
derived in \S~4 is used.
For the subsample of non-detections, we estimated the upper limits
for the source count rates using the local RASS exposure at the
 optical positions
of the objects and assume  less than 12 source photons
from the objects.
This assumption is based on the fact that
almost all sources which were found with a detection likelihood
less than $10$ have less than 12 source photons.
For a source within a background of strong diffuse X-ray emission
or in the vicinity of another strong X-ray source,
this estimate for the upper limit might not be correct.
However, we expect that such cases are rather rare
and that they do not affect our statistical results.
Furthermore, for sources showing abnormally weak X-ray emission
compared to their optical brightness, we have visually checked the RASS
source images to avoid possible inappropriate estimates of upper limits.
For more than 100 objects which were found with
detection likelihoods $\ge 7$ ($\ga 3~\sigma$) in the RASS
but lower than our threshold of 10 and which were thus not regarded
as detections, we used the measured count rates as upper limits.
In the optical band, the $B$ magnitudes are taken from the VV93 catalogue.
We apply an extinction correction
using the fitted relation between reddening and Galactic column
density for a constant gas-to-dust ratio (Burstein \& Heiles 1978).
Magnitude corrections due to emission lines were made
as described in Marshall \eta (1983).
The luminosities at $2500\AA$ are calculated by assuming
an optical spectral index $\alpha = 0.5$.

In \S~5.1 we study the luminosity correlation between the optical and
X-ray bands. Then we investigate the distribution of the X-ray-to-optical
luminosity ratio \alpox in \S~5.2, and a study of its dependence on
\z and \lo is presented in \S~5.3.

\subsection{Luminosity correlations}   

The proper treatment of redshift effects is one of the biggest
difficulties in the investigation of luminosity correlations.
As our sample is relatively large,
we are able to minimize these effects by using
objects within a narrow redshift bin.
For the analysis we consider objects with $z < 1$ only,
since at $z \ga 1$
the low fraction of X-ray detections of our sample ($\la 5\%$, see Fig.~3)
can hardly yield reliable results.
We split the redshift range $z=0.1-1.0$ into 9 bins with a bin size of 0.1.
The changes of the luminosity distance square within a redshift  bin 
 of $\Delta z = 0.1$ are about factors of 4 and 1.2
at $z=0.15$ and $z=0.95$, respectively,
which is small compared to the corresponding range of luminosities 
of more than one order of magnitude.
The objects within a single redshift  bin can thus be regarded as being
at approximately the same distance.

For the subsample of each  \z bin we tested the \lx -- \lo correlation,
using a Spearman correlation test incorporating upper limits
(using the survival analysis software ASURV; Rev 1.3, LaValley \eta 1992).
The probability level for rejecting the ``no correlation'' hypothesis, $P_r$, ranges
from $0.47\times 10^{-2}$ to $0.11\times 10^{-5}$,
with a median of $0.84\times 10^{-3}$ for all nine subsamples.
The weakest correlation ($P_r=0.47\times 10^{-2}$)
occurs in the lowest redshift bin, $z=$~0.1--0.2,
where the intrinsic luminosity range is rather small
($\log~l_o=29.8 - 30.6~ $\mlume).
We thus confirm the correlation between \lo and \lx
being an intrinsic property for most objects of the
quasar population at redshifts $z=0.1-1$,
rather than a distance or a selection effect.
 However, despite the presence of a significant correlation
 the data points show a rather large scatter around the 
 regression line.
Part of the scatter in the low-\z ($\la 0.5$) bins,
which show relatively large $P_r$ values, comes from a few objects
showing rather weak X-ray luminosities compared to their 
optical luminosities ($\alpha_{ox} \ga 2.0$),
deviating from the rest of the subsample.
Excluding objects with $\alpha_{ox} > 2.0$  (for details see \S~5.2)
yielded generally reduced $P_r$ values except 
in the $0.1 \leq z < 0.2$  bin.

As an example, we plot in Fig.~8
the \lx -- \lo relationship for the subsample in the $z=0.3-0.4$ bin,
in which the upper limits of the X-ray luminosities for the non-detections
(47 out of 103 objects) are indicated by arrows.
The dotted lines indicate constant luminosity ratios
of $\alpha_{ox} = 1.2, 1.5$ and 1.9, respectively.
It shows that for most of the objects
\lx  roughly scales with \lo over about two orders of magnitudes,
though with large scatter.
Outliers are two objects with $\alpha_{ox} > 1.9$,
showing much weaker X-ray emission by a factor of $\sim 10$
than expected for objects with $\alpha_{ox}=1.5$,
which is roughly the mean \alpox value.

\begin{figure}
\psfig{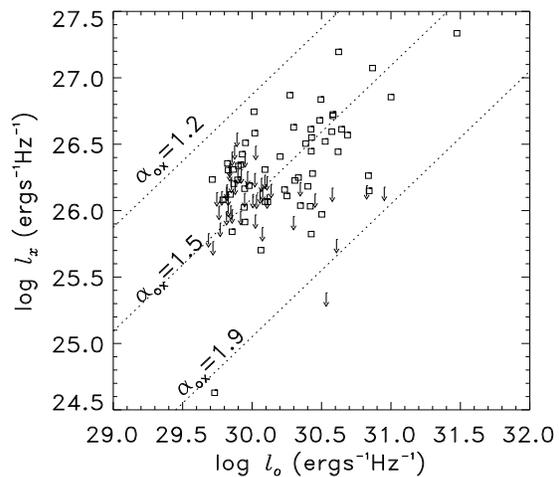}
\caption[]{The relation between the monochromatic \lx and \lo
for the subsample in the redshift range $0.3 < z < 0.4$.
Detections are indicated by
squares and X-ray upper limits by arrows.
The dotted lines indicate constant X-ray-to-optical luminosity ratios
of $\alpha_{ox} = 1.2, 1.5$, and 1.9, respectively. }
\end{figure}

Various methods of regression analyses were applied to determine
the slope  of the correlation $\l_x \sim l_o^e$, 
taking into account upper limits, errors in one or both variables and/or 
assuming an intrinsic dispersion of the data.
We obtained slopes ranging from $\sim$ 0.7 to $\sim$ 1.2
depending on the applied mathematical method and the analyzed data set 
(see the discussion of this problem in Babu \& Feigelson 1996).
For the analysis of the subsamples in different redshift bins the 
typical $1~\sigma$ errors are $\Delta e \sim 0.2$, due to the small 
sample sizes.
   
\begin{table*}
\caption{Regression analysis for the total quasar  sample}
    \begin{flushleft} \begin{tabular}{cllcl}
    \hline\noalign{\smallskip}
     Method~$^{a)}$ & $\Delta x$ & $\Delta y$& $e$ & Source code \\
    \noalign{\smallskip}  \hline \noalign{\smallskip}
     OLS & no error & no error & 0.75$\pm$0.03 &  IDL/Num.Rec. \\
     OLS & no error & $\Delta l_x = 1\sigma $ & 0.814$\pm$0.002 &  IDL/Num.Rec. \\
     OLS & no error & $\Delta l_x = 0.3$ & 0.75$\pm$0.01 &  IDL/Num.Rec. \\ 
     ODR & $\Delta$ m = 0.2$^{b)}$ &  $\Delta l_x = 1\sigma^{c)}$ & 1.00$\pm$0.07 &
      FV, $\sigma=0$,~fixed \\
     ODR  & $\Delta$ m = 0.2 & $\Delta l_x = 0.3$ & 0.91$\pm$0.07 &  FV, $\sigma=0$,
                      ~fixed  \\
     ODR  & $\Delta$ m = 0.2 & $\Delta l_x = 1\sigma$ & 1.16$\pm$0.03 &  FV, $\sigma=0.23
       \pm 0.02$ \\
     ODR  & $\Delta$ m = 0.2 & $\Delta l_x = 0.3$ & 1.15$\pm$0.03 &  FV, $\sigma=0.22
       \pm 0.02$ \\
    \noalign{\smallskip}  \hline
    \end{tabular}  \end{flushleft}
\smallskip \noindent \\
$^{a)}$ OLS: ordinary least squares regression, ODR: orthogonal distance regression  \\
$^{b)}$ Errors in optical luminosities are calculated assuming $\Delta m_v = 0.2$
      magnitudes \\
$^{c)}$  Errors in X-ray luminosities are either the $1~\sigma$ statistical errors or,
   for illustrative purposes, 
   an assumed error of 30\% to account for possible systematic uncertainties.
\end{table*}
  
For illustrative purposes we present in Table 4  the results
of various methods for the determination of $e$ for the total
sample of  the detected radio-quiet quasars (objects with \alpox $ >$ 2 were excluded).
The results obviously depend on the applied method and on the errors
of the individual data points.
For a discussion of the methods see paper~I.
It has been argued (La Franca \eta 1995) that ODR methods yield 
more reliable results for data with errors in both variables.
 Further, we note that the modified 
orthogonal distance regression method (FV, Fasano \& Vio 1988) shows that there is 
significant dispersion in the data indicating that
there is not the same strict one-to-one correlation between the two 
variables for all of the objects.
 An explanation for this result could be that the quasar sample does not form
a homogeneous group of objects or that, for example, the orientation of the quasar
to the line of sight affects the correlation given that  X-rays and
 optical flux do not show the same angular dependence.

\subsection{The \alpox distribution}

To find the intrinsic distribution of the X-ray loudness \alpox
we used a maximum-likelihood method, the ``detections and bounds'' or DB method
 developed by Avni \eta (1980, 1995) which takes 
 into account both detections and upper limits
(lower limits for \alpoxe) for the non-detections.
This method gives reliable results only for randomly distributed censored data.
Although the X-ray flux limits for the non-detections
are set by the limiting sensitivity of the RASS
and are thus relatively uniform,
the limits on X-ray-to-optical luminosity ratios are, however,
roughly randomly distributed.
We derive the normalized distribution of \alpox in 20 bins  
from $\alpha_{ox}=0.4$ to 2.4 
($f_i(\alpha_{ox}), i=1,\cdots,20$;  $\sum_{i=1}^{20}f_i(\alpha_{ox})=1$),
in a non-parametric way.
Such an analysis has the advantage of being able to determine the
shape of the distribution without further a\,priori assumptions.
We restrict the analysis to a subsample of objects with
optical $B$-magnitudes  brighter
than 18.5 in order to include a considerable fraction of detections
(see Fig.~2 for the magnitude dependent X-ray detection rate).
In fact, for objects much fainter than $B \sim 18.5$,
the X-ray fluxes expected from the typical range of \alpox for radio-quiet quasars
are generally below the flux limit of the RASS.
This would lead to highly biased
and thus meaningless lower limits of $\alpha_{ox}$ for the non-detections.
The subsample thus comprises 377 detections and 989 non-detections,
with a detection rate of $\sim 30\%$.

Fig.~9 shows the normalized histogram of $\alpha_{ox}$ for the detections
(open line) and the best estimated ``unbiased'' distribution function
derived from the DB method taking into
account the lower limits of $\alpha_{ox}$ for non-detections (shaded).
For the 377 X-ray detected objects (when 
the method reduces to the conventional sample statistics) 
we find a mean \malpox = $1.54 \pm 0.01$, 
a standard deviation $\sigma \sim 0.17\pm 0.01$
($\sigma^2=\sum_{i=1}^{20}f_i (\alpha_{ox,i}-\mbox{\malpoxe})^2$),
and a skewness $= 0.29^{+0.34}_{-0.40}$
($\sum_{i=1}^{20}f_i (\alpha_{ox,i}-\mbox{\malpoxe})^3/\sigma^3$);
while the resulting maximum likelihood distribution
including the non-detections yields a mean \malpox = $1.65 \pm 0.01$,
a standard deviation $\sigma=0.19\pm 0.01$, and a skewness $=0.43^{+0.39}_{-0.27}$,
with a tail towards higher $\alpha_{ox}$
(the uncertainties are at 68\% confidence level for one interesting parameter;
see Avni \eta 1980, 1995).
The results also indicate a large dispersion in the \alpox distribution,
as could already be inferred in \S~3.2 from the detection of objects in the RASS
over a relatively wide range of optical magnitudes.

\begin{figure}
\psfig{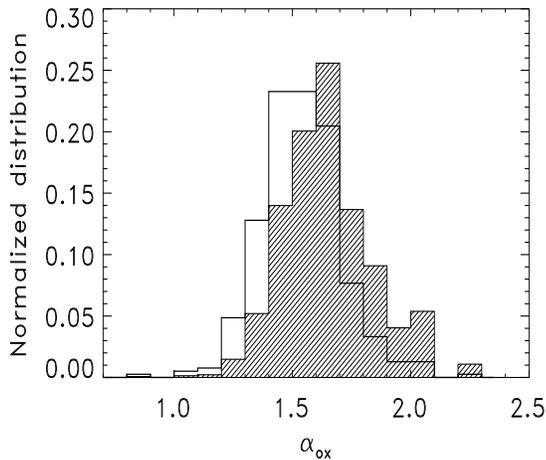}
\caption[]{Normalized distribution
of the X-ray-to-optical luminosity ratios \alpox   for
the magnitude-limited subsample of objects with  $B \leq 18.5$.
The open line is for the objects detected in X-rays,
and the shaded area represents the
best estimate distribution taking into account non-detections using the
maximum likelihood DB method (see the text).}
\end{figure}

Of particular interest is a small number
of objects showing very weak X-ray emission at 2~keV compared to their 
optical luminosity, as mentioned in \S~3.2 and \S~5.1.
In Fig.~10 we plot  the distribution of \alpox for the detected  
$z < 0.5$ objects in the magnitude-limited subsample,
in order to avoid the ``normal'' high \alpox values expected
for large \z for a flux-limited sample.
It shows more clearly a tail towards higher \alpoxe,
with six objects having  $\alpha_{ox} > 2.0$.
Further, there are two objects which were not detected
but must have values of $\alpha_{ox} > 1.95$.
These objects show highly reduced
X-ray emission by more than a factor of 30
compared to the luminosities of the
bulk of the detected objects at the same optical luminosity.
These low redshift sources  were detected 
in pointed observations only,
except PG 0844+349 (at a high state) and PG 1001+05,
which were also seen in the RASS.
In fact, the non-detection of these objects in the RASS
is the major cause of the reduced detection rate
at bright magnitudes ($B \sim 15$) as shown in Fig.~2.
Amongst the 20 nearby quasars with $z < 0.1$
in our optically selected sample,
there are 3 of these ``X-ray quiet'' objects. The existence of 
 ``X-ray quiet'' objects was also claimed by Laor \eta (1997) on the basis of
3 out of 23 objects, of which 2 are radio-quiet and included in our sample.
It is not clear whether these objects are the X-ray quiet extremes of
a continuous distribution or a distinct class of objects.
We note, however, that X-ray variability might play a role in these objects;
for example, PG 0844+349 appears to be ``X-ray quiet'' only in its low state.
In addition, the possibility for some of them being previously unknown BAL objects,
which remain undetected in soft X-rays (Green \eta 1995, Green \& Mathur 1996),
cannot be ruled out as well.

\begin{figure}
\psfig{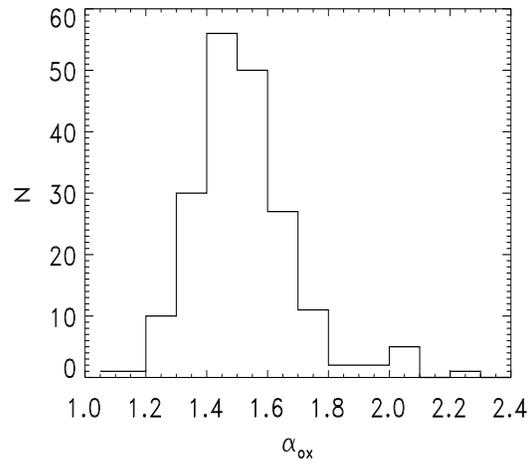}
\caption[]{Histogram of the \alpox distribution for
objects with  $z < 0.5$ from the magnitude-limited subsample.
It shows a X-ray quiet tail towards higher \alpoxe.}
\end{figure}
   
Excluding the 6 objects with  $\alpha_{ox} > 2.0$
and the two objects with lower limits of
$\alpha_{ox} > 1.95$, we obtained for the remaining 
371 detected objects and 987 non-detections a slightly reduced mean
and standard deviation (\malpox $= 1.63 \pm 0.01$, $\sigma = 0.16\pm 0.01$),
and a reduced best estimate for the skewness = $-0.12^{+0.48}_{-0.19}$, 
i.e., compatible with a symmetric distribution around the mean.

The such determined mean \alpox is consistent,
within the mutual errors, with the values found in previous studies
(for example, \malpox$=1.57 \pm 0.15$
for the LBQS sample with a limiting magnitude at $B\sim 19$, Green \eta 1995).
The marginally higher best estimate arises from the brighter cut-off (18.5th mag)
of our magnitude-limited subsample,
as the observed \alpox appears to increase for high-\lo objects.
The detection limit of the X-ray observations
and the incompleteness of a sample
affect the exact value of \malpox as well.
Finally, we used the relatively well determined individual spectral
properties for many of the sources for the determination of the
monochromatic X-ray luminosities instead of an overall
spectral slope, which  reduces the systematic uncertainties.

\subsection{Dependence of \alpox on redshift and optical luminosity}

We test the dependence of \alpox on \z and \lo
by applying both the non-parametric and parametric forms of the DB method
to the magnitude-limited subsample.
We excluded from the subsample the six objects with  $\alpha_{ox} > 2.0$ 
and the two  non-detections with lower limits \alpox $> 1.9$
from the following analysis,
for they show extremely weak X-ray emission
compared to the bulk of the objects.

We first divided the subsample into six redshift bins, as defined in Table~5,
in which the number of objects, the obtained \malpox and standard deviations $\sigma$
in each bin are listed.
Beyond a redshift of $z = 0.5$, the \malpox are consistent with one another
within their mutual $1~\sigma$ errors;
whereas the $z < 0.2$ bin shows a significantly lower \malpoxe.
The probability that the $z < 0.2$ bin has the same \alpox distribution
as the higher redshift bins is less than $ P < 0.01$.
In fact, the mean \alpox in the $z < 0.2$ bin (\malpox$=1.49\pm0.02$) is well determined,
in the sense that all the objects were detected and the DB method is reduced to
the conventional evaluation of the mean.
The \malpox for the 126 detections in the $0.2 \leq z < 0.5$ bin
is \malpox $\sim 1.53$, whereas the 90 undetected objects
have an average lower limit of \alpox of $\sim 1.57$.
It is not clear from the analysis
whether \malpox depends primarily on $z$ or on \loe,
given the presence of a  correlation between $z$ and \lo
for our magnitude-limited sample.
We will investigate below whether at low redshifts ($z < 0.5$),
\malpox  depends  on both, \z and \loe.
The standard deviations of the \alpox distributions
are consistent for all the $z$-bins.

\begin{table}
\caption{Maximum-likelihood $\alpha_{ox}$ distribution in redshift bins}
    \begin{flushleft} \begin{tabular}{lcccc}
    \hline\hline\noalign{\smallskip}
      redshift bins & $N_d^{a)}$ & $N_b^{b)}$ & \malpox & $\sigma$ \\
    \noalign{\smallskip} \hline \noalign{\smallskip}
    $      z <    0.2$ &58 & 0 & $1.49^{+0.02}_{-0.02}$ & $0.14$ \\
    $0.2 \leq z < 0.5$ &126& 90& $1.60^{+0.03}_{-0.02}$ & $0.19$ \\
    $0.5 \leq z < 1.0$ &83 &161& $1.63^{+0.04}_{-0.03}$ & $0.18$ \\
    $1.0 \leq z < 1.5$ &39 &208& $1.69^{+0.02}_{-0.03}$ & $0.14$ \\
    $1.5 \leq z < 2.0$ &29 &228& $1.68^{+0.06}_{-0.02}$ & $0.13$ \\
    $  z \ge 2.0     $ &36 &299& $1.69^{+0.03}_{-0.03}$ & $0.15$ \\
    \noalign{\smallskip}  \hline
    \end{tabular}  \end{flushleft}
Note: errors are at 68\% confidence for 1 interesting parameter \\
a) number of objects detected in X-rays \\
b) number of objects with lower limits in \alpox
\end{table}

To investigate the $\alpha_{ox}-l_o$ dependence
we divided the subsample into six  $\log~l_o$ bins, as defined in Table~6.
The luminosity intervals were chosen such 
that the number of detections in a bin are roughly comparable.
The derived \malpox and 68\% errors in all $\log~l_o$ bins
are listed in column~2 (case A) of Table~6,
and are shown in Fig.~11,
where the data points (only detections for clarity) are plotted.
For comparison the ``X-ray quiet'' objects ($\alpha_{ox} > 2$, filled circles)
are indicated as well.
The mean \alpox shows a general increase with $\log~l_o$; however, 
for low luminosities, $\log~l_o  \la 30.5$,  the \malpox are consistent with
being independent of $\log~l_o$.
This result is even more pronounced  in  case B (column~3 of Table~6),
where the $z < 0.2$ objects with their significantly lower \malpox were excluded
to reduce the potential \z dependences.
The inclusion of the ``X-ray quiet'' objects does not affect the results.
A similar trend of the  $\alpha_{ox} - l_o$ relation 
has been noted by Avni \eta (1995) in a study of the $Einstein$ quasar sample,
and it can be inferred as well from the data of
the radio-loud ROSAT quasars in paper~I (Fig.~17)
and those of the RASS-LBQS sample in Green \eta (1995, Fig.~4b).

\begin{table*}
\caption{Maximum-likelihood results for the mean \malpox as a function of
$\log~l_o$ and $z$}
    \begin{flushleft} \begin{tabular}{lccccccc}
    \hline \hline\noalign{\smallskip}
    $\log~l_o$  & \multicolumn{2}{c}{all redshifts} & & \multicolumn{4}{c}{redshift ranges} \\
    \cline{2-3} \cline{5-8}
    \mlum &case A$^{a)}$& case B$^{b)}$ & & $z < 0.2$ & $0.2 \leq z < 0.5 $ &
    $0.5 \leq z < 1.0$ & $z \ge 1.0$\\
    \noalign{\smallskip}  \hline \noalign{\smallskip}
      $\log~l_o < 30            $ & $1.53^{+0.03}_{-0.03}$ & $1.58^{+0.06}_{-0.04}$ & & $1.47^{+0.04}_{-0.03}$ & $1.58^{+0.06}_{-0.04}$ &         --             &           --           \\
      $30   \leq \log~l_o < 30.2$ & $1.55^{+0.03}_{-0.02}$ & $1.57^{+0.07}_{-0.03}$ & & $1.51^{+0.03}_{-0.03}$ & $1.55^{+0.06}_{-0.03}$ &         --             &           --           \\
      $30.2 \leq \log~l_o < 30.5$ & $1.56^{+0.02}_{-0.02}$ & $1.58^{+0.05}_{-0.03}$ & & $1.52^{+0.05}_{-0.05}$ & $1.59^{+0.07}_{-0.04}$ & $1.57^{+0.04}_{-0.03}$ &           --           \\
      $30.5 \leq \log~l_o < 31.0$ & $1.64^{+0.03}_{-0.02}$ & $1.64^{+0.03}_{-0.02}$ & &       --               & $1.70^{+0.10}_{-0.06}$ & $1.60^{+0.03}_{-0.02}$ & $1.66^{+0.18}_{-0.05}$ \\
      $31.0 \leq \log~l_o < 31.5$ & $1.65^{+0.02}_{-0.02}$ & $1.65^{+0.02}_{-0.02}$ & &       --               &      --                & $1.75^{+0.14}_{-0.09}$ & $1.64^{+0.02}_{-0.02}$ \\
$\log~l_o \geq 31.5       $ & $1.73^{+0.02}_{-0.02}$ & $1.73^{+0.02}_{-0.02}$ & &       --               &      --                &     --                 & $1.73^{+0.02}_{-0.02}$ \\
    \noalign{\smallskip}  \hline
    \end{tabular}  \end{flushleft}
Note: errors are at 68\% confidence for 1 interesting parameter \\
a) using the magnitude-limited subsample \\
b) the same as a) but with $z < 0.2$ objects excluded
\label{table:tab5}
\end{table*}

\begin{figure}
\psfig{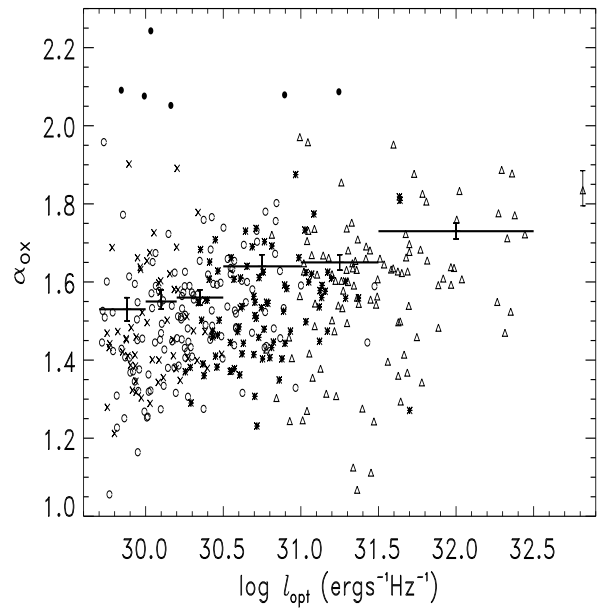}
\caption[]{The \alpox -- $\log~l_o$ relation for the
subsample of $B \le 18.5$; only detections are plotted.
The error bar at log~$l_o \sim 32.8$ indicates the typical $1~\sigma$ error of the
measured $\alpha_{ox}$.
The \malpox and 68\% errors, obtained by taking into account non-detections,
are plotted for each log~\lo bin.
Plot symbols are:
crosses for $z < 0.2$ objects;
open circles for $0.2 \le z < 0.5$ objects;
asterisks for $0.5 \le z < 1.0$ objects;
triangles for $z \ge 1.0$ objects;
filled circles for the ``X-ray quiet'' objects ($\alpha_{ox} > 2$).}
\end{figure}

The sample is large enough  to allow an attempt for a  non-parametric test for
a joint \alpox dependence on both, \lo and \z, in the \lo -- \z plane.
We subdivided the objects in each luminosity interval
into four redshift ranges (distinguished by various symbols in Fig.~11),
and list the obtained \malpox in columns 4--7 of Table~\ref{table:tab5}, respectively.
An empty entry means that there are no or too few objects in the bin 
to yield reliable results.
Although the binned data are too sparse to give  statistically convincing
results, a few  conclusions can be inferred from the table:
(a) In a given $\log~l_o$ range the \malpox 
     at different \z are consistent with each other
    within their mutual $1~\sigma$ errors
    except for $\log~l_o < 30$,  where the $z < 0.2$ objects have a
    significantly lower ($2~\sigma$ confidence) \malpox 
    than  in the $0.2\le  z < 0.5$ range.
    This implies that the low \malpox in the lowest-\z bin ($z< 0.2$)
    found above arises from a true $z$-dependence at low redshifts,
    rather than as a consequence of a luminosity dependence.
    This redshift dependence  is significant only for $\log~l_o < 30$ objects.
(b)  For any redshift range the largest \malpox  is found in 
    the highest $\log~l_o$ bin  
    though the significance is generally not high ($< 1.5~\sigma$),
    whereas the \malpox in the other \lo bins of that redshift range
    are consistent with each other.
    This could imply that the \alpox-- \lo dependence,
    found for the sample as a whole
    holds for every narrow redshift range.

In an alternative approach, we used the parametric form of the DB method
(Avni \& Tananbaum 1986, Wilkes \eta 1994)
to fit a dependence of the form
\begin{equation}
\mbox{\malpox}=A_z[\tau(z)-0.5]+A_o(\log~l_o-30.5) + A
\end{equation}
to the data,
where $\tau(z)$ is the look-back time in units of the Hubble time
($\tau(z)=1-(1+z)^{-3/2}$ for $q_0=0.5$). The fit assumes that 
the residuals of the \alpox have a Gaussian distribution.
In Fig.~12 we show  the joint 90\% confidence contours for
$A_z$ and $A_o$ (two interesting parameters)
for the subsample (solid line),
with the best estimates of $A_z=-0.016$ and $A_o=0.11$.
The results are consistent with those from the above non-parametric method and
with previous studies
(Kriss \& Canizares 1985, Avni \& Tananbaum 1986, Wilkes \eta 1994),
showing a dependence of \malpox on $\log~l_o$, but not on redshift $z$.
Excluding objects with $z < 0.2$ and $\log~l_o < 30.0$, as mentioned above,
leads to similar results (dotted line).
Considering only low-$z$ objects ($z < 0.5$), however, the contours (dashed line) show
a slight dependence on $z$
(an increase of \alpox by $\sim 0.12$ from the present epoch to half the Hubble time,
i.e.,  $z \sim 0.6$, for the best fitted  $A_z=0.25$), at 90\% confidence level,
in addition to the dependence on $\log~l_o$.
The dependence of \malpox on \z at low redshifts ($z < 0.5$),
found in a non-parametric way above, is confirmed.

\begin{figure}
\psfig{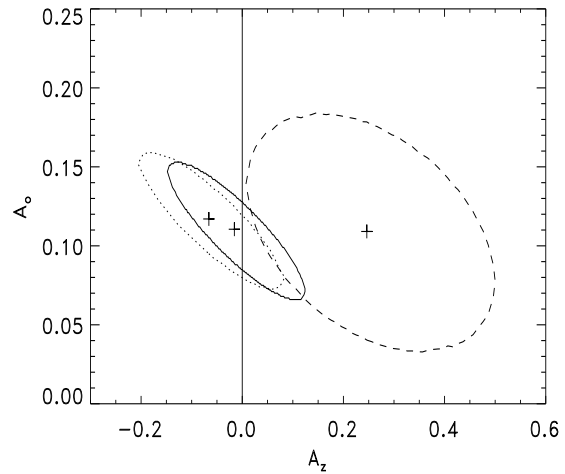}
\caption[]{The 90\% confidence contours for the two parameters $A_z$ and $A_o$
of the \alpox -- $z$, $\log~l_o$ dependence for the magnitude-limited
subsample of $B \le 18.5$, where the ``X-ray quiet'' objects ($\alpha_{ox}>2$)
have been excluded. Solid line: all objects; dotted line: all objects, but sources at 
$z < 0.2$ and $\log(l_o) <$ 30.0 excluded ; dashed line: objects at $z < 0.5$ only.  }
\end{figure}

In summary, our analyses show
a dependence of \alpox on $\log~l_o$ rather than on $z$.
Two new features can be noted:
up to $z \la 0.5$, radio-quiet quasars,
mostly the low-\lo objects ($\log~l_o < 30$),
show a slight increase in \malpox with $z$.
Secondly, the \alpox -- log~\lo dependence is complex and 
the increase of \alpox with \lo is  significant only for $\log~l_o \ga 30.5$.

It should be noted that the $ z < 0.2$ objects have on average the lowest optical 
luminosities among the quasar population but higher than those of Seyfert~I galaxies
which have a mean \malpox $\sim$ 1.25 (Kriss \& Canizares 1985).
Thus, our low-{\it z}, low-\lo sample might be contaminated by objects with 
quasar luminosities but still having Sy~I - type properties. These objects would
not be detected at higher redshifts and could thus lead to the observed 
\malpox dependence with redshift. 
   
Finally, 
the increase of \malpox with $\log~l_o$, inferred from the statistical analysis of
the data, can not necessarily be interpreted as an 
 intrinsic relationship of the quasar population.
In paper~I we have demonstrated that the 
apparent $\alpha_{ox}$ -- $\log~l_o$ correlation
can emerge even for an intrinsically uncorrelated sample,
due to the inherent large dispersion in the scaling law between \lx and \lo
and the flux limitations of the observations, leading to a rhomboidal shaped
phase space diagram (Fig. 19 of paper~I). A detailed Monte Carlo simulation,
addressing the question whether an underlying truly physical  correlation 
can be extracted from the 
data in the presence of these biases  will be presented elsewhere
(Yuan \eta 1997).

\section{Summary}

We have presented the X-ray spectra and flux densities for
846 radio-quiet quasars detected by ROSAT in both
the All-Sky Survey and in pointed observations.
This is the largest sample
of X-ray detected radio-quiet quasars published so far,
with many of the objects ($\sim$ 70\%) being detected in X-rays for the first time.
We studied the broad band properties of the population by
using a sub-sample of optically selected, radio-quiet quasars
from this compilation, which comprises 644 detections.
The large number of objects enables us to
investigate their properties in a wide range of parameter space
with high statistical significance.
Our major results are the following:

1. For radio-quiet quasars we found a systematically
lower and much faster decreasing X-ray detection probability
in the RASS towards faint optical magnitudes and high redshifts
than for radio-loud objects (paper~I).
The on average higher X-ray luminosity of radio-loud quasars
can account for these differences qualitatively.

2. A large fraction of the quasars,
observed more than once in pointed observations,
shows moderate flux variations, mostly by less than a factor of two.
Extreme flux variations (by a factor of six)
were found for only one radio-quiet object (PG~0844+349).

3. We confirm that radio-quiet quasars have in general steeper
soft X-ray spectra (\mGammag = $2.58 \pm 0.05$ for  $z < 0.5$)
than radio-loud objects
($\Delta\Gamma \sim 0.4 $ and $\Delta\Gamma \sim 0.3$, compared to 
flat- and steep-spectrum radio quasars, respectively) and that
this spectral difference persists at $z > 2$
with $\Delta\Gamma \sim 0.6$ compared to the flat-spectrum quasars (paper~I).
It is also confirmed that the soft X-ray spectra in the ROSAT band
are systematically steeper than those in the harder $Einstein$ IPC band 
by $\Delta\Gamma \sim 0.5$.

4. A spectral flattening with redshift is also confirmed for radio-quiet objects,
which can be described by a linear relation of the form 
$\mbox{\mGammag}= 2.63(\pm 0.04) - 0.20(\pm 0.04) \times z$, for $z \la 2.0$.
As for the class of radio-loud quasars,
the spectral slopes appear to be independent of redshift beyond $z \sim 2$.
A decrease of the dispersion of the distribution of spectral slopes with redshift
(from $\sigma=0.38^{+0.05}_{-0.04}$ for $z < 0.5$ to
      $\sigma=0.26^{+0.20}_{-0.12}$ for $z > 2.5$)
is also inferred, though the significance is not high.
These results are consistent with a composite X-ray spectral model including
an additional steep soft X-ray component.

5. The spectral slope of $\Gamma \sim 2.23^{+0.16}_{-0.19}$ found 
  for the ROSAT radio-quiet quasars at high redshifts ($z > 2.5$),
is consistent, within the errors,
with those found for nearby quasars in the medium energy X-ray band
(2--10~keV) of $\Gamma \sim 2.0$.
This implies that X-ray spectral evolution is not important in radio-quiet quasars.
The steep spectra of radio-quiet quasars seen at $z > 2.5$
strengthen the claims that
they cannot be the dominant contributors to the cosmic X-ray background.

6. The data are, in a statistical sense,
consistent with no or only little excess absorption
at all redshifts for radio-quiet quasars,
though the significance of the result beyond $z \sim 2$ is not high.

7. By dividing the sample into narrow redshift bins ($\Delta z = 0.1$) for $z < 1$,
the correlation between the X-ray luminosity and the luminosity at $2500\AA$ is
explicitly tested and proved to be an intrinsic property
of the quasar population, though a large scatter is present in the correlation.
The slopes of the regression line, obtained from  different  mathematical methods,
cover a wide range, $0.7 \la e \la 1.2$.
The existence of an intrinsic non-zero dispersion in the correlation between 
 X-ray and optical luminosities suggest the existence of an additional 
``variable'', not accounted for in the analysis. 
One possibility would be a non-uniform, orientation dependent attenuation of the 
fluxes in the different energy bands.  

8. In accordance with previous studies, the X-ray loudness \alpox seems to
   be independent of redshift but to correlate with $\log~l_o$.
   The \alpox -- log~\lo dependence is complex:
   the increase of \alpox with \lo is significant only for $\log~l_o \ga 30.5$
   and  the low-\lo  radio-quiet quasars show a slight
    increase of  \malpox with $z$ up to $z \la 0.5$.
   However,  this correlation might be primarily caused by luminosity
  selection biases and by the intrinsic dispersion of the data,
  as already discussed in paper~I for radio loud quasars.
  A detailed study of these effects by Monte Carlo simulations
   will be presented elsewhere (Yuan \eta 1997).
   
9. Highly reduced X-ray emission, by more than a factor of $\sim 30$ compared to
    the bulk of the objects, is  
    found in a few objects in the high-\alpox tail of the \alpox distribution.
   There are eight  such ``X-ray quiet'' objects below $z < 0.5$, six  
   detected by ROSAT and two upper limits, having $\alpha_{ox} > 2$.
   The reason for the low X-ray emission  of these objects is unknown;
    we note, however, that X-ray variability might play a role.
 For example, the highly variable object PG 0844+349 is classified 
  as ``X-ray quiet'' only in its low state.

\vskip 0.4cm
\begin{acknowledgements}
The ROSAT project is supported by the Bundesministerium f\"ur
Bildung, Wissenschaft, Forschung und Technologie (BMBF) and
the Max-Planck-Gesellschaft.
We thank our colleagues from the ROSAT group for their support,
and M. Veron-Cetty for supplying a machine readable
form of their catalogue.
This research has made use of the NASA/IPAC Extragalactic Data Base
(NED) which is operated by the Jet Propulsion Laboratory,
California Institute of Technology, under contract with the National
Aeronautics and Space Administration.
\end{acknowledgements}

\end{document}